\documentclass[11pt,twoside]{article}
\usepackage{asp2010}
\resetcounters

\aspvolauthor{Marie Treyer, Ted Wyder, Don Neill, Mark Seibert, Janice Lee, eds.}
\aspvoltitle{``UP: Have Observations Revealed a Variable Upper End of the Initial Mass Function?''}
\aspcpryear{20XX}
\aspvolume{4XX}

\begin{document}

\title{Measuring the Upper End of the Initial Mass Function with Supernovae}
\author{James D. Neill$^1$
\affil{$^1$California Institute of Technology, 1200 East California Blvd., Pasadena, CA  91125}}

\begin{abstract}
Supernovae arise from progenitor stars occupying the upper end of the
initial mass function.  Their extreme brightness allows individual massive
stars to be detected at cosmic distances, lending supernovae great
potential as tracers of the upper end of the IMF and its evolution.
Exploiting this potential requires progress in many areas of supernova
science.  These include understanding the progenitor masses that produce
various types of supernovae and accurately characterizing the supernova
outburst and the environment in which it was produced.  I present some
preliminary work identifying the environmental conditions that produce the
most luminous supernovae, believed to arise from stars with masses greater
than 100 $M_{\odot}$.  I illustrate that the presence of these extreme 
supernovae in small star-forming dwarfs can be used to test our 
understanding of the upper end of the IMF.
\end{abstract}

\section{Motivation}

Supernovae (SNe) are the spectacular end result of massive star evolution.
Current estimates of the lower mass limit for SN progenitors are near 8
$M_{\odot}$ \citep[e.g.,][]{Smartt:09:1409}, making them extremely relevant
to studies of the upper end of the initial mass function (IMF).  The fact
that SNe are now being detected beyond $z = 2$ \citep{Cooke:09:237} adds
the potential to use SNe as tracers of IMF evolution.  Their usefulness
depends on how well their progenitors can be understood and how well the
observations of SNe can constrain the distribution of progenitor masses.
In addition, we must be able to characterize the host galaxies of SNe and
connect them to our general understanding of star formation in galaxies at
various stages of their evolution.

\section{SN Wish List}

Let us take an idealized view of SNe and list what they have to offer for
studies of the upper-IMF.  For a given galaxy with a given star formation
history (SFH) or current star formation rate (SFR), we would like to know:
what is the mass of the largest star that can be produced?  This is a
fiducial point in the IMF that calibrates the relationship betwee the IMF
and the host SFR \citep[e.g.,][]{Weidner:10:275}.  The fact that such
massive stars are bright is offset by their very short lives, making them
difficult to observe.  The SN explosion marks these massive stars in events
detectable across cosmic distances.  Provided we can decode the progenitor
mass and measure the host galaxy SFH/SFR, we can directly calibrate this
important fiducial point in a system's IMF, and trace its evolution
potentially out beyond $z=2$.

If we know the masses of not just the most massive SN progenitors, but a
whole range of progenitors down to the 8 $M_{\odot}$ limit, we can measure
the distribution of massive stars from 8 to over 100 $M_{\odot}$ as a
function of host SFR/SFH and mass and, while we are at it, metallicity.
Assuming we have stellar evolution models for these massive stars that will
provide us with their lifetimes, we can then look into the evolution of
clusters with greater accuracy, revealing how energy and processed metals
are re-introduced into the birth cloud of our idealized SN progenitors.

Now that we have imagined this ideal universe in which understanding the
upper end of the IMF rests only on the collection of enough SNe and the
corollary host data to sample the distributions well enough, let us now
look at the reality of SN progenitors and how close we can get to this
idealized situation.

\section{Reality}

\begin{table}[h]
\begin{tabular}{lllcr}
{\bf Type}& {\bf Mechanism}& {\bf Character}& {\bf Mass ($M_{\odot}$)} &
{\bf Progenitor}\\
\hline\\
Ia	& Thermonuclear & mass-transfering binary	& 3 -- 8	& CO WD \\
II(P,L,n) & Core-Collapse & H in absorption (emission)	& 7 -- 25	& R/BSG \\
IIb	& Core-Collapse & weak H simlar to He	& 15 -- 25	& (late) WN \\
Ib	& Core-Collapse & all H removed		& 25 -- 40	& (early) WN \\
Ic	& Core-Collapse & all H, He removed	& 40 -- 80	& WC/WO \\
IIn-lum	& Core-Collapse & narrow H emission	& 80 -- 150	& LBV \\
Ipec	& pulsational pair? & all H, He removed	& 80 -- 150	& W-R? \\
I-PP	& pulsational pair & Fe-group but no H, He	& $>$100	& pop III?
\end{tabular}
\caption{SN Types\label{tab_sn_types}}
\end{table}

I will briefly outline the menagerie of SN types and what we know about
their progenitors.  I have summarized the basic properties of SNe of
various types in Table~1 after the scheme presented in \citet{Yam:07:372}.
SNe are divided roughly into two categories depending on the presence (type
II) or absence (type I) of hydrogen, either in emission or absorption, in
their spectra.  When we attribute physical processes to the SN explosions
it becomes clear that a more rational division is between those
explosions caused by the collapse of a massive stellar core
(core-collapse, or CC SNe) and those explosions caused by the
thermonuclear burning of a carbon-oxygen white dwarf (thermonuclear, or
type Ia SNe).  We will not consider the thermonuclear SNe, since we know
they arise in binary systems \citep[for a review, see][]{Livio:01:334} and
it is unlikely that their progenitors arise from stars more massive than 8
$M_{\odot}$, otherwise they would produce CC SNe.

Within the CC SN types, there appears to be a natural progression of
atmospheric depletion due to mass loss of varying efficiency.  The least
depleted are the type II SNe that show hydrogen lines.  These SNe are
further sub-divided by light curve shape (II-P for `plateau' and II-L for
`linear decline') and the presence of narrow emission lines (II-n),
indicating the interaction of the SN ejecta with surrounding circum-stellar
material.  The type IIb SNe represent the next level of depletion in which
helium lines are as strong or stronger than the hydrogen lines.  Type Ib
SNe occur when the atmospheric depletion is so complete that only helium
lines are seen.  The most depleted objects are the type Ic SNe, showing
neither hydrogen nor helium lines in their spectra.  Some of these depleted
objects have cores so massive that they undergo a pulsational
pair-production instability which eventually produces a very luminous
explosion.  These are the pair instability SNe (PISNe) that have been
theoretically predicted for decades \citep{Barkat:67:379, Bond:84:825,
Heger:02:532, Waldman:08:1103}, but only recently confirmed observationally
\citep{Yam:09:624,Quimby:09}.

The cause of the mass loss that depletes the hydrogen and helium is some
combination of low surface gravity, stellar winds, and binary interaction.
This leads to significant uncertainty when estimating progenitor initial
masses.  We know that metallicity plays a role in determining the
efficiency of wind-driven stellar mass loss, but there is significant
uncertainty in this theory.  The ugly reality is that the type of the SN
derives from some unknown combination of metallicity, binary interaction,
and intitial stellar mass, clouding our vision of SNe as ideal probes of
the upper IMF.

As of today, the best constrained SN progenitors are the least depleted
ones near the lower mass limit.  \citet{Smartt:09:1409} report
pre-explosion multi-band photometry of at least seven SNe II-P and limits
on another 13.  These result in initial mass estimates for the progenitors
in the range from $\sim7 - 25 M_{\odot}$.  This is a good beginning, but
this sample represents a small fraction of the population of SN II SNe.
There is no way to calculate an initial mass distribution from this small
sample.

\section{The Dawning Time-Domain Era}

\begin{table}[h]
\begin{tabular}{lrrlccc}
	{\bf Survey}& {\bf Tel.}& {\bf FOV} & {\bf Filters} & {\bf Cadence}& {\bf
Depth/day} &
{\bf Coverage}\\
\hline\\
PTF & 1.2m & 7$^{\circ}$& gR & 1-5day & $\sim$21 mag & 8000$^{2\circ}$/yr\\
PanSTARRS & 1.8m         & 7$^{\circ}$& grizy & 4day & 22.3(y), 25.1(i) &
3$\pi$ - 30,000$^{2\circ}$\\
CSS & 0.7m & 8$^{\circ}$& unflt. & 1day & $\sim$20 mag & 1200$^{2\circ}$/dy\\
SkyMapper & 1.3m & 6$^{\circ}$& uvgriz & 4hrs-1yr & $\sim$22(g) & 2$\pi$ -
26,000$^{2\circ}$\\
LSST & 6.5m & 10$^{\circ}$& ugrizy3 & 3day & 24.5(r) & 3300$^{2\circ}$/dy \\
\end{tabular}
\caption{Current and Future Transient Surveys\label{tab_surveys}}
\end{table}

There is no doubt that progress on the upper IMF requires improvements in
the theory of high-mass stellar evolution.  More precise and divers models
of SN explosions will also help in deciphering progenitor characteristics
from the spectra of SNe in outburst.  The engine capable of driving
meaningful theoretical progress is a more comprehensive and representative
data set covering the full diversity of SN explosions.  We are on the verge
of a quantum jump in the quality and the quantity of observations of SNe in
outburst with wide-field transient surveys like PanSTARRS
\citep{Hodapp:04:636}, PTF \citep{Law:09:1395}, CSS \citep{Drake:09:870},
SkyMapper \citep{Keller:07:1} already online and others such as LSST
\citep{Ivezic:08:2366} coming online in the future.

These surveys, summarized in Table~\ref{tab_surveys}, take advantage of
wide-field detectors on robotically controlled telescopes to produce large
area time-domain surveys that are already having an impact on our
understanding of SN demographics.  Old SN surveys used narrow-field
detectors and were neccessarily host-targeted, producing biases tied to the
luminosity and mass of hosting galaxies.  The new areal searches are not
only removing these biases, but are discovering new types of SNe.

These surveys have the potential to vastly improve our measurements of SN
demographics and produce accurate distributions of types and rates of each
type.  With the proper followup, these rates can be corellated with host
galaxy SFR, stellar mass, and average metallicity.  They have already shown
their value by discovering a new type of extremely luminous SN, but it is
likely that other rarities will be discovered as well.

\section{Tantalizing Highlight: Extreme SNe in Extreme Hosts}

The first addition to the SN menagerie made by these areal SN surveys
already shows the power of comprehensive surveys to illuminate regions of
parameter space that have been in darkness up to now.  Because of the host
targeted bias, necessitated by narrow field searches, the SNe in faint
hosts have been passed over.  It turns out that the most luminous SNe on
record are typically found in faint, small dwarf galaxies \citep{Young:09,
Yam:09:624, Quimby:09}.  These luminous SNe (LSNe) include very luminous
explosions showing narrow emission lines that help to power the later light
curve and are known as SNe IIn-lum.  There is another set of LSNe that show
no H in their spectra, but have unusual light curves and spectra when
compared to any other SN type.  These we call SNe Ipec.

Recent analysis shows that two LSNe (SN1999as and SN2007bi) show strong
evidence of having been produced as the result of the pulsational pair
instability process \citep{Yam:09:624}, hence we label them Ic-PP.  Their
light curves are consistent with models of extreme LSNe that produce
$>10^{52}$ ergs of total energy implying the radioactive decay of a large
amount ($>4M_{\odot}$) of $^{56}$Ni.  Their spectra show the results of
this $^{56}$Ni decay by having a high abundance of Fe-group elements
\citep{Yam:09:624}.  These signatures require a very massive initial mass
for the progenitor star, upwards of 150 $M_{\odot}$.  There is also
evidence that the progenitor of SN2007bi was a single star
\citep{Yam:09:624}.  It is possible that some of the SNe Ipec that have
similar properties may also be PISNe, but they may not be as extreme.

\begin{figure}
\includegraphics[angle=90.,scale=0.6,viewport=24 54 528 608]{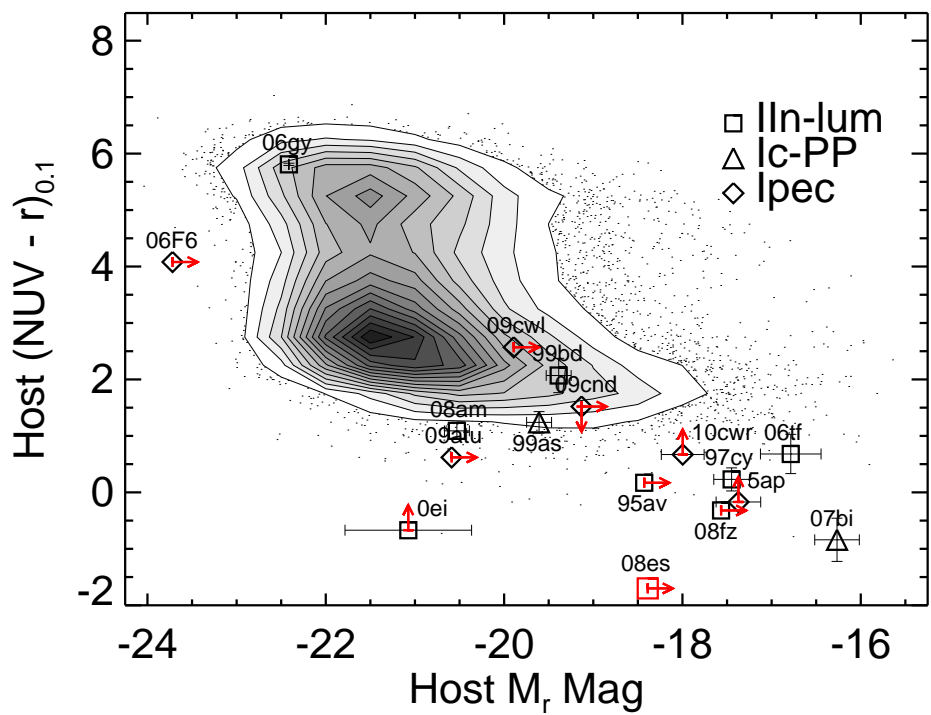}
\caption{Galaxy $NUV-r$ versus $M_r$ CMD with extreme SNe indicated
\citep{Neill:10}.  SN types are indicated by the symbols with IIn-lum SNe
indicated by squares, Ic-PP SNe (PISNe) indicated by triangles, and Ipec
SNe indicated by diamonds.  Right pointing arrows indicate which half-plane
the particular LSN host is limited to by the UV-optical photometry, upward
pointing arrows indicate that the $NUV-r$ color must be redder than the
symbol, while the double arrows for PTF09cnd indicate the quarter-plane
that its host is limited to.  The contours represent the galaxy density per
bin (0.5 by 0.5 mag) of the comparison sample which is derived from the
{\it GALEX} -- SDSS cross-match presented in \citet{Wyder:07:293}.  The
darkest contour indicates a density of 1056 galaxies per bin, while the 
lightest bin is at 132 galaxies per bin.  Below the minimum contour
galaxies are plotted individually as small points.  The LSN hosts appear to
favor hosts on the blue side of the blue cloud, toward the low-luminosity
end.
\label{figcmd} }
\end{figure}

We analyzed the hosts of seventeen LSNe \citep{Neill:10} using {\it GALEX}
\citep{Martin:05:L1} and SDSS \citep{York:00:1579} images to derive their
UV-optical colors and estimate their stellar masses ($M_*$) and specific
SFR ($sSFR=SFR/M_*$). We compare them to a larger sample of nearby galaxies
\citep{Wyder:07:293} to illustrate how extreme the hosts are and to
determine what phase of galaxy evolution might be demarcated by LSNe in
such small galaxies.  While the incompleteness in the comparison sample at
the faint end prevents a determination of the relative space density of
LSNe producers, one can see (Figure~\ref{figcmd}) that LSNe appear to
prefer small star-bursting dwarfs.

The LSN hosts have low total SFR, but high sSFR due to their low masses.
The progenitor of one of the LSNe (SN2007bi) may exceed 150 $M_{\odot}$.
Any theory of the IMF that posits an upper limit to the most massive star
produced based on current SF must account for these cases, which may be
considered limiting.  To illustrate this, consider Figure~4 of
\citet{Altenburg:07:1550}.  Here we see that for SFRs near $10^{-1}
M_{\odot}$ yr$^{-1}$, a 150 $M_{\odot}$ star is just barely consistent.
This is probably good news for such theories, because these object are
indeed rare.  Currently the errors on both the host SFR and the progenitor
initial mass are large, but as these shrink, the LSNe and their hosts could
drive a re-calibration of IMF theories that predict the mass of the most
massive star produced based on current star formation.

\section{Conclusions}

While the reality of using SNe to map out the entire upper end of the IMF
appears to be a long way off, there is cause for hope for the future.
Large area surveys will produce unbiased samples of SNe, allowing us to
test and improve theories of SN progenitors and their production.  Even
though we may never be able to map out the distribution of high mass stars
with SNe, it does seem likely that LSNe may arise from some of the most
massive stars known, providing a crucial calibration point in IMF theory.
The fact that LSNe appear to favor small galaxies with low mass, low
intrinsic SFR, but high sSFR makes them even more useful to advancing IMF
theory.

\acknowledgements 

GALEX (Galaxy Evolution Explorer) is a NASA Small Explorer, launched in
2003 April. We gratefully acknowledge NASA's support for construction,
operation, and science analysis for the GALEX mission, developed in
cooperation with the Centre National d'Etudes Spatiales of France and the
Korean Ministry of Science and Technology.

Funding for the SDSS and SDSS-II has been provided by the Alfred P.
Sloan Foundation, the Participating Institutions, the National Science
Foundation, the U.S. Department of Energy, the National Aeronautics and
Space Administration, the Japanese Monbukagakusho, the Max Planck
Society, and the Higher Education Funding Council for England. The SDSS
Web Site is http://www.sdss.org/.

This research has made use of the NASA/IPAC Extragalactic Database (NED)
which is operated by the Jet Propulsion Laboratory, California Institute of
Technology, under contract with the National Aeronautics and Space
Administration.


\begin{thebibliography}{}
\expandafter\ifx\csname natexlab\endcsname\relax\def\natexlab#1{#1}\fi
\expandafter\ifx\csname url\endcsname\relax
  \def\url#1{\texttt{#1}}\fi
\expandafter\ifx\csname urlprefix\endcsname\relax\def\urlprefix{URL }\fi
\providecommand{\eprint}[2][]{\url{#2}}

\bibitem[{Barkat et~al.(1967)Barkat, Rakavy, \& Sack}]{Barkat:67:379}
Barkat, Z., Rakavy, G., \& Sack, N. 1967, Physical Review Letters, 18, 379

\bibitem[{Bond et~al.(1984)Bond, Arnett, \& Carr}]{Bond:84:825}
Bond, J.~R., Arnett, W.~D., \& Carr, B.~J. 1984, \apj, 280, 825

\bibitem[{Cooke et~al.(2009)Cooke, Sullivan, Barton, Bullock, Carlberg,
  Gal-Yam, \& Tollerud}]{Cooke:09:237}
Cooke, J., Sullivan, M., Barton, E.~J., Bullock, J.~S., Carlberg, R.~G.,
  Gal-Yam, A., \& Tollerud, E. 2009, \nat, 460, 237

\bibitem[{Drake et~al.(2009)Drake, Djorgovski, Mahabal, Beshore, Larson,
  Graham, Williams, Christensen et~al.}]{Drake:09:870}
Drake, A.~J., Djorgovski, S.~G., Mahabal, A., Beshore, E., Larson, S., Graham,
  M.~J., Williams, R., Christensen, E., et~al. 2009, \apj, 696, 870

\bibitem[{Gal-Yam et~al.(2007)Gal-Yam, Leonard, Fox, Cenko, Soderberg, Moon,
  Sand, Li et~al.}]{Yam:07:372}
Gal-Yam, A., Leonard, D.~C., Fox, D.~B., Cenko, S.~B., Soderberg, A.~M., Moon,
  D.-S., Sand, D.~J., Li, W., et~al. 2007, \apj, 656, 372

\bibitem[{Gal-Yam et~al.(2009)Gal-Yam, Mazzali, Ofek, Nugent, Kulkarni,
  Kasliwal, Quimby, Filippenko et~al.}]{Yam:09:624}
Gal-Yam, A., Mazzali, P., Ofek, E.~O., Nugent, P.~E., Kulkarni, S.~R.,
  Kasliwal, M.~M., Quimby, R.~M., Filippenko, A.~V., et~al. 2009, \nat, 462,
  624

\bibitem[{Heger \& Woosley(2002)}]{Heger:02:532}
Heger, A., \& Woosley, S.~E. 2002, \apj, 567, 532

\bibitem[{Hodapp et~al.(2004)Hodapp, Kaiser, Aussel, Burgett, Chambers, Chun,
  Dombeck, Douglas et~al.}]{Hodapp:04:636}
Hodapp, K.~W., Kaiser, N., Aussel, H., Burgett, W., Chambers, K.~C., Chun, M.,
  Dombeck, T., Douglas, A., et~al. 2004, Astronomische Nachrichten, 325, 636

\bibitem[{Ivezic et~al.(2008)Ivezic, Tyson, Allsman, Andrew, Angel, \& for~the
  LSST.~Collaboration}]{Ivezic:08:2366}
Ivezic, Z., Tyson, J.~A., Allsman, R., Andrew, J., Angel, R., \& for~the
  LSST.~Collaboration 2008, arXiv:astro-ph, 0805, 2366

\bibitem[{Keller et~al.(2007)Keller, Schmidt, Bessell, Conroy, Francis,
  Granlund, Kowald, Oates et~al.}]{Keller:07:1}
Keller, S.~C., Schmidt, B.~P., Bessell, M.~S., Conroy, P.~G., Francis, P.,
  Granlund, A., Kowald, E., Oates, A.~P., et~al. 2007, Publications of the
  Astronomical Society of Australia, 24, 1

\bibitem[{Law et~al.(2009)Law, Kulkarni, Dekany, Ofek, Quimby, Nugent, Surace,
  Grillmair et~al.}]{Law:09:1395}
Law, N.~M., Kulkarni, S.~R., Dekany, R.~G., Ofek, E.~O., Quimby, R.~M., Nugent,
  P.~E., Surace, J., Grillmair, C.~C., et~al. 2009, \pasp, 121, 1395

\bibitem[{Livio(2001)}]{Livio:01:334}
Livio, M. 2001, In: Supernovae and gamma-ray bursts: the greatest explosions
  since the Big Bang. Proceedings of the Space Telescope Science Institute
  Symposium, 334

\bibitem[{Martin et~al.(2005)Martin, Fanson, Schiminovich, Morrissey, Friedman,
  Barlow, Conrow, Grange et~al.}]{Martin:05:L1}
Martin, D.~C., Fanson, J., Schiminovich, D., Morrissey, P., Friedman, P.~G.,
  Barlow, T.~A., Conrow, T., Grange, R., et~al. 2005, \apj, 619, L1

\bibitem[{Neill et~al.(2010)Neill, Sullivan, Gal-Yam, Quimby, Ofek, Wyder,
  Howell, Nugent et~al.}]{Neill:10}
Neill, J.~D., Sullivan, M., Gal-Yam, A., Quimby, R., Ofek, E., Wyder, T.~K.,
  Howell, D.~A., Nugent, P.~E., et~al. 2010, \apj, accepted, astro-ph.CO.
  \eprint{1011.3512v2}

\bibitem[{Pflamm-Altenburg et~al.(2007)Pflamm-Altenburg, Weidner, \&
  Kroupa}]{Altenburg:07:1550}
Pflamm-Altenburg, J., Weidner, C., \& Kroupa, P. 2007, \apj, 671, 1550

\bibitem[{Quimby et~al.(2009)Quimby, Kulkarni, Kasliwal, Gal-Yam, Arcavi,
  Sullivan, Nugent, Thomas et~al.}]{Quimby:09}
Quimby, R.~M., Kulkarni, S.~R., Kasliwal, M.~M., Gal-Yam, A., Arcavi, I.,
  Sullivan, M., Nugent, P., Thomas, R., et~al. 2009, arXiv:astro-ph,
  astro-ph.CO. \eprint{0910.0059v1}

\bibitem[{Smartt et~al.(2009)Smartt, Eldridge, Crockett, \&
  Maund}]{Smartt:09:1409}
Smartt, S.~J., Eldridge, J.~J., Crockett, R.~M., \& Maund, J.~R. 2009, \mnras,
  395, 1409

\bibitem[{Waldman(2008)}]{Waldman:08:1103}
Waldman, R. 2008, \apj, 685, 1103

\bibitem[{Weidner et~al.(2010)Weidner, Kroupa, \& Bonnell}]{Weidner:10:275}
Weidner, C., Kroupa, P., \& Bonnell, I. A.~D. 2010, \mnras, 401, 275

\bibitem[{Wyder et~al.(2007)Wyder, Martin, Schiminovich, Seibert, Budav{\'a}ri,
  Treyer, Barlow, Forster et~al.}]{Wyder:07:293}
Wyder, T.~K., Martin, D.~C., Schiminovich, D., Seibert, M., Budav{\'a}ri, T.,
  Treyer, M.~A., Barlow, T.~A., Forster, K., et~al. 2007, \apjs, 173, 293

\bibitem[{York et~al.(2000)York, Adelman, Anderson, Anderson, Annis, Bahcall,
  Bakken, Barkhouser et~al.}]{York:00:1579}
York, D.~G., Adelman, J., Anderson, J.~E., Anderson, S.~F., Annis, J., Bahcall,
  N.~A., Bakken, J.~A., Barkhouser, R., et~al. 2000, \aj, 120, 1579

\bibitem[{Young et~al.(2009)Young, Smartt, Valenti, Pastorello, Benetti, Benn,
  Bersier, Botticella et~al.}]{Young:09}
Young, D.~R., Smartt, S.~J., Valenti, S., Pastorello, A., Benetti, S., Benn,
  C.~R., Bersier, D., Botticella, M.~T., et~al. 2009, arXiv:astro-ph,
  astro-ph.CO. \eprint{0910.2248v1}

\end{thebibliography}

\end{document}